\documentclass[a4paper,12pt]{article}
\usepackage{amsmath}
\usepackage{amssymb}
\usepackage{amsfonts}
\usepackage{fleqn}
\usepackage[footnotesize]{caption}
\usepackage{mathrsfs}

\def\I{\mathrm{i}}
\def\SU{\text{SU}}

\def\U{\text{U}}

\def\eps{\epsilon}
\def\epsb{\bar{\epsilon}}

\def\epstensor{\varepsilon}

\def\Fla{\theta}
\def\Flb{\bar{\theta}}

\def\<{\left\langle}
\def\>{\right\rangle}
\def\ChargeC{\mathrm{C}}

\allowdisplaybreaks[2]

\begin{document}

\bibliographystyle{OurBibTeX}

\begin{titlepage}

 \vspace*{-15mm}
\begin{flushright}
SHEP/0510\\
UMD-PP-05-042
\end{flushright}
\vspace*{5mm}

\begin{center}
{
\sffamily
\LARGE
Quark-Lepton Complementarity \\[2mm] in Unified Theories
}
\\[8mm]
S.~Antusch\footnote{E-mail: \texttt{santusch@hep.phys.soton.ac.uk}}$^{(a)}$,
S.~F.~King\footnote{E-mail: \texttt{sfk@hep.phys.soton.ac.uk}}$^{(a)}$
R.~N.~Mohapatra\footnote{E-mail: \texttt{rmohapat@physics.umd.edu}}$^{(b)}$
\\[3mm]
{\small\it $^{(a)}$
Department of Physics and Astronomy,
University of Southampton,\\
Southampton, SO17 1BJ, U.K.
}\\[1mm]
{\small\it $^{(b)}$
Department of Physics, University of Maryland, \\ College Park, MD 20742, USA
}

\end{center}
\vspace*{0.75cm}

\begin{abstract}

\noindent 
As pointed out by many authors, recent observations are consistent with 
an intriguing relation between the Cabibbo angle $\theta_{\mathrm{C}}$
and the solar neutrino mixing angle $\theta_{12}$, namely 
$\theta_{12}\simeq \pi/4 - \theta_{\mathrm{C}}$.
Such quark-lepton complementarity (QLC) may be a signal
of an underlying quark-lepton unification at short distances. We
discuss possible ways to realize this relation in realistic quark-lepton
unification theories by identifying a minimal set of operators
that lead to QLC while remaining consistent with other known
data. The purpose of this paper is to present 
the first elements of a unified model at the GUT scale 
capable of predicting the QLC relation.
A generic prediction of our proposed class of models
is the new relation for the lepton mixing angle
$\theta_{13} \simeq \theta_{\mathrm{C}}$, which
allows these models to be confirmed or excluded by
the current generation of neutrino oscillation experiments.

\end{abstract}

\end{titlepage}
\newpage
\setcounter{footnote}{0}

\section{Introduction}
The unusual nature of neutrino mixings as compared to the quark
mixings has inspired a large amount of speculation regarding 
symmetries in the quark-lepton world as well as other kinds of new
physics beyond the standard model \cite{King:2003jb}. 
A particularly intriguing clue is the observation that the
leptonic mixing angle $\theta_{12}$ and the Cabibbo angle
$\theta_{\mathrm{C}}$ (which provides the quark 1-2 mixing) seem to satisfy a
relation,
\begin{eqnarray}\label{Eq:QLCrelation}
\theta_{12}+\theta_{\mathrm{C}}\approx\frac{\pi}{4}.
\end{eqnarray}

In the context of inverted hierarchy models it was
observed some time ago that the predictions for the neutrino mixing angles
of $\theta^\nu_{12}=\pi/4$, $\theta^\nu_{13}=0$, 
may receive corrections from the charged
lepton mixing angle of order the Cabibbo angle, 
$\theta^e_{12}\approx \theta_{\mathrm{C}}$, resulting in 
the lepton mixing angle $\theta^\nu_{12}$ being in the LMA MSW
range, with $\theta^\nu_{13}$ close to its current experimental
limit (see e.g.\ \cite{King:2000ce,King:2002nf,Frampton:2004ud}).
However the possibility of a precise relation such as in 
Eq.~(\ref{Eq:QLCrelation}) was not considered.

While it is quite possible that Eq.~(\ref{Eq:QLCrelation}) 
is purely accidental, recent
papers \cite{Raidal:2004iw,Minakata:2004xt,QLCliterature} have speculated that this could be a signal
of some high scale quark-lepton unification. 
Indeed the structure of the
Standard Model (SM) itself may be said to already provide some
indirect hints of such a high scale unification, for example from the
universality of weak currents, and the
SU(2)$_\mathrm{L}$ assignment of fermions.
However the question of how to extend the SM in such a way as
to lead to a rather precise QLC relation in Eq.~(\ref{Eq:QLCrelation}) 
is far from clear, and this is the subject of 
this paper.

The general strategy we shall follow here is to start from 
an inverted hierarchy structure of the neutrino mass matrix,
which predicts $\theta^\nu_{12}=\pi/4$, $\theta^\nu_{13}=0$, 
then look for corrections from the charged lepton sector
which can not only give acceptable corrections to the 
lepton mixing angles as suggested in \cite{King:2000ce,King:2002nf}, 
but can accurately give rise to the QLC
relation in Eq.~(\ref{Eq:QLCrelation}). In order to 
achieve this, it seems necessary to relate the
charged lepton mixing angle $\theta^e_{12}$ to the Cabibbo angle
$\theta_{\mathrm{C}}$, and this in turn only seems possible
within the framework of a quark-lepton symmetry.
As we shall see, it is highly non-trivial to obtain the 
QLC relation in Eq.~(\ref{Eq:QLCrelation}) for several reasons.
On the other hand, if such a relation is in fact realized, then this
potentially tells us a great deal about the detailed structure of the
unified theory, as we shall see. A smoking gun signature of the class
of models that we propose here is the additional QLC relation:
\begin{eqnarray}\label{Eq:auxQLCrelation}
\theta_{13}\approx\theta_{\mathrm{C}},
\end{eqnarray}
which enables the approach presented here to be verified or excluded
by the present generation of neutrino experiments.

In a recent paper 
\cite{mohfram} a model based on the 
Pati-Salam gauge group
SU(2)$_\mathrm{L}\times$ SU(2)$_\mathrm{R} \times$ SU(4)$_\mathrm{C}$ 
that unifies quarks and
leptons \cite{ps} was proposed in an attempt to realize the QLC relation.
It was pointed out that an inverted hierarchy model based on the
$L_\mathrm{e}-L_\mu -L_\tau$ symmetry imposed on single bi-doublet
Pati-Salam model could be a good starting point for realizing QLC
in a natural gauge theory framework.
It was realized in \cite{mohfram} that obtaining realistic models
for both the quark and lepton sector is highly nontrivial in such
models and inevitably leads to nontrivial
constraints on the nature of the theory. In particular, Ref.~\cite{mohfram}
could lead only to
a modified form of QLC (see discussion below) as well as the
scale of quark lepton unification to be at the multi-TeV scale.
Furthermore, the relation between the Cabibbo angle and the solar mixing
angle obey a complementarity relation that is different from that in 
Eq.~(\ref{Eq:QLCrelation})
(although it is consistent with observations at 2$\sigma$ level).

It is the goal of this paper to discuss models for QLC that are more
in the spirit of grand unified theories (GUTs)
so that quarks and leptons unify at the
GUT scale of $10^{16}$ GeV. Our strategy will be to construct from the
see-saw mechanism a form of neutrino mass matrix consistent with an 
inverted hierarchy, using the approach 
of \cite{King:2000ce,King:2002nf}.
Note that in this approach it is possible that the effective neutrino 
mass matrix has a form which is invariant under
$L_\mathrm{e}-L_\mu-L_\tau$ without ever imposing this symmetry in 
the high energy theory. In fact we shall instead rely on a different
$U(1)_\mathrm{X}$ family symmetry as playing a leading role in
achieving this structure. In order to achieve QLC, it is necessary to 
achieve this within the framework of quark-lepton unification, 
which we shall assume here to be provided by a high energy Pati-Salam
theory, broken near the unification scale.
The presence of the Pati-Salam gauge group is necessary to relate the
charged lepton mixing angles to the down-quark mixing angles,
which in turn dominate the contribution to the Cabibbo angle,
allowing the QLC relation to emerge, albeit in a non-trivial way.
Our approach in this paper is to list explicitly 
a set of higher dimensional operators in a
minimal supersymmetric Pati-Salam type theory that after symmetry
breaking leads to the required form for the mass matrices. We have not
attempted to seek a set of discrete symmetries that would enable us to
generate the right set of higher dimensional operators, nor have we 
addressed the question of the orderings of the index contractions
of these operators, since both questions would take us well beyond
the scope of this paper. However we have identified 
a $U(1)_\mathrm{X}$ family symmetry which controls the leading
operators, which would facilitate the construction of 
a future theory from the building blocks that we present here.
The purpose of this paper is therefore to present 
the first elements of a unified model at the GUT scale 
capable of predicting the QLC relation.

This paper has been organized as follows: in Sec.~2, we discuss
generic problems with realizing QLC relation in models where the
$SU(4)_\mathrm{C}$ quark-lepton unification scale is close to the GUT
scale; in Sec.~3, we review a way of generating
inverted hierarchy (different from that in Ref.~\cite{mohfram}
which uses $L_\mathrm{e}-L_\mu -L_\tau$ symmetry); in Sec.~4, we present a
the set of operators that enables us to realize the inverted hierarchy
for neutrinos and eventually, helps us realize the QLC relation.
In Appendix A we specify our conventions, and
in Appendix B we briefly discuss an alternative class of 
QLC models with a lopsided charged lepton mass matrix.

\section{Challenges for QLC}\label{Sec:TextureExample}

We shall see that it is not enough to simply postulate
a Pati-Salam symmetry and an inverted hierarchical effective
neutrino mass matrix: there are a number of additional hurdles
one must clear before the QLC relation will emerge.
The purpose of this section is to outline the main challenges
facing QLC. These challenges can be regarded as a chain of logic,
that must remain unbroken at each stage in order to achieve the
desired result of QLC. We shall find that there are non-trivial
obstacles facing this approach.

The basic starting point of our approach 
is to assume that the low energy neutrino mass matrix $m_{\mathrm{LL}}$
accurately predicts $\theta_{12}^{\nu} = \pi/4$.
We have already observed that this
is possible if the neutrino mass spectrum is inverted, with a
Majorana parity for $m_1$ and $m_2$.
In the see-saw scenario, $m_{\mathrm{LL}}$ is given via the see-saw formula
\begin{eqnarray}
m_{\mathrm{LL}} = v_\mathrm{u}^2 Y_\nu M_\mathrm{RR}^{-1}Y_\nu^T  \; ,
\end{eqnarray}
where $Y_\nu$ is the
neutrino Yukawa matrix,
$M_\mathrm{RR}$ the mass matrix of the right-handed neutrinos and
$v_\mathrm{u} = \< h^0_\mathrm{u} \>$
is the vev which leads to masses for the up-type quarks.
We will discuss the construction of see-saw models with inverted hierarchy
in Sec.~\ref{Sec:Inverted}. For the time being let us assume that nearly
maximal neutrino mixing $\theta_{12}^{\nu}$ 
is generated from $m_{\mathrm{LL}}$, which is then subsequently
corrected by charged lepton mixing angles in forming the 
final leptonic mixing angle $\theta_{12}$.
Clearly these corrections coming from the charged lepton
sector must have a rather special form, if QLC is to emerge.

In our conventions, which are
specified in the appendix, the MNS matrix is given by
$U_\mathrm{MNS}=U_{e_\mathrm{L}} U_{\nu_\mathrm{L}}^\dagger$. 
If we parameterize
$U_{e_\mathrm{L}}$ and $U_{\nu_\mathrm{L}}$ in standard parameterization,
we obtain (neglecting the very small mixings $\theta_{23}^\mathrm{e}$ and
$\theta_{13}^\mathrm{e}$):
\begin{eqnarray}
U_\mathrm{MNS} \approx R^{\mathrm{e}\dagger}_{12}
R^\nu_{23}U^\nu_{13}R^\nu_{12}P_0^\nu \; .
\end{eqnarray}
In general, 
small contributions from the charged lepton mixing angles
$Y_\mathrm{e}$ lead to the following corrections
to the mixings from the neutrino sector \cite{King:2002nf}:
\begin{subequations}\label{Eq:YeContributions}
\begin{eqnarray}
s_{23} &=& s_{23}^\nu - \theta_{23}^\mathrm{e} c_{23}^\nu \; ,\\
\theta_{13} &=& \theta_{13}^\nu - \theta_{13}^\mathrm{e} c_{23}^\nu
               + \theta_{12}^\mathrm{e} s_{23}^\nu \; ,\\
\label{Eq:ChargedLContrToT12} s_{12} &=& s_{12}^\nu + \theta_{23}^\mathrm{e} s_{12}^\nu
               + \theta_{13}^\mathrm{e} c_{12}^\nu s_{23}^\nu
           - \theta_{12}^\mathrm{e} c_{23}^\nu c_{12}^\nu   \; ,
\end{eqnarray}
\end{subequations}
where we have neglected phases for simplicity.
Assuming $\theta_{23}^\nu \approx \pi/4$, we obtain
\begin{subequations}\label{Eq:YeContrTexture1}
\begin{eqnarray}
s_{23} &\approx& s_{23}^\nu \; ,\\
\label{Eq:YeContrTexture1_T13_Example}\theta_{13} &\approx& \theta_{13}^\nu
               +\tfrac{1}{\sqrt{2}} \theta_{12}^\mathrm{e}\; ,\\
\label{Eq:ChargedLContrToT12_Example}\label{Eq:YeContrTexture1_T12_Example}
s_{12} &\approx& \tfrac{1}{\sqrt{2}}
           - \tfrac{1}{2} \theta_{12}^\mathrm{e}
         \;\;  \rightarrow \;\;
                \theta_{12} \;\:\approx \;\:\tfrac{\pi}{4} - \tfrac{1}{\sqrt{2}}\theta_{12}^\mathrm{e}
         \; .
\end{eqnarray}
\end{subequations}
We immediately see one problem for realizing QLC:
with maximal atmospheric mixing $\theta_{23}^\nu = \pi/4$ from
$m_{\mathrm{LL}}$, shifting the rotation $R^{\mathrm{e}\dagger}_{12}$ 
to the right
side of $R^\nu_{23}$ induces a factor of $1/\sqrt{2}$
for the correction to $\theta_{12}$ coming from the charged leptons.
Therefore the charged lepton mixing angle
$\theta_{12}^\mathrm{e}$ needs to be about $\sqrt{2}$ larger than
$\theta_C$ in order to compensate for this.
Note Eq.~(\ref{Eq:ChargedLContrToT12_Example}) implies that the correction to 
$\theta_{13}$ from the charged lepton sector is also equal to
$1/\sqrt{2}$ times the charged lepton angle $\theta_{12}^\mathrm{e}$.

In our approach it is necessary that 
that the Cabibbo angle be dominated by the down quark 1-2 mixing angle,
$\theta_C \approx \theta^d_{12}$, with the up-quark 1-2 mixing angle
$\theta^u_{12}$ being essentially irrelevant.
This is because we shall subsequently use quark-lepton symmetry to relate 
$\theta_{12}^\mathrm{e}$ to $\theta_{12}^\mathrm{d}$.
Consider a symmetric texture for the quark Yukawa matrices
of the form\footnote{For a detailed discussion of the this texture
see, e.g., \cite{Roberts:2001zy}.}
\begin{eqnarray}
\begin{array}{ll}
Y_{\mathrm{u}}=
\begin{pmatrix}
0        & b'\eps^3 & c'\eps^4 \\
b'\eps^3  & \eps^2 & a'\eps^2 \\
c'\eps^4  & a'\eps^2 & 1 \\
\end{pmatrix} y_\mathrm{t} \;,
&
Y_\mathrm{d}=
\begin{pmatrix}
0        & b\epsb^3 & c\epsb^4 \\
b\epsb^3  & \epsb^2 & a\epsb^2 \\
c\epsb^4  & a\epsb^2 & 1 \\
\end{pmatrix} y_\mathrm{b}\;.
\end{array}
\end{eqnarray}
With parameters $\eps$ and $\epsb$ determined from the quark masses of the
second and third generation at the unification scale $M_{X} \approx 2 \cdot
10^{16}\:\mbox{GeV}$,
\begin{eqnarray}
\frac{m_\mathrm{c}}{m_\mathrm{t}} = \eps^2 \; \rightarrow \eps \approx 0.05 \; ,
\quad
\frac{m_\mathrm{s}}{m_\mathrm{b}} = \epsb^2 \; \rightarrow \epsb \approx 0.15 \;,
\end{eqnarray}
it allows to fit the quark data. Constraints from the CKM matrix can be
satisfied by taking e.g.\
\begin{eqnarray}
b \approx 1.5 \; , \quad
c \approx 3 \; , \quad
a \approx 1.3 \; , \quad
|b'| \approx 1 \; , \quad
\mbox{Arg}(b') \approx \tfrac{\pi}{2} \; .
\end{eqnarray}
The other parameters, $a'$ and $c'$ are rather unconstrained. The
texture zero in the (1,1)-entry of $Y_{\mathrm{u}}$ and $Y_\mathrm{d}$ leads to
the famous GST relation \cite{GST} which connects the quark masses and the Cabibbo angle
\begin{eqnarray}
\theta_{\mathrm{C}} = |V_{us}| = \lambda =
\left|\sqrt{\frac{m_\mathrm{d}}{m_\mathrm{s}}} -
e^{-\I \phi}\sqrt{\frac{m_\mathrm{u}}{m_\mathrm{c}}}\right| ,
\end{eqnarray}
with $\phi = \mbox{Arg}(b')$.
For the above parameter set it is seen to be the case that 
the Cabibbo angle is numerically approximately given from the 
down quark sector only, 
\begin{equation}
\theta_C \approx \theta^d_{12}\approx 
\sqrt{\frac{m_\mathrm{d}}{m_\mathrm{s}}}
\end{equation}
which satisfies our necessary condition of QLC. 

Having argued that it is plausible that 
$\theta_C \approx \theta^d_{12}$, 
the next link in the chain of logic leading to QLC is to relate the 
down quark mixing angle $\theta^d_{12}$ to the charged lepton mixing
angle $\theta^e_{12}$. The well known observation is that 
the above texture can accommodate the charged lepton data as well, if a
Clebsch factor (Georgi-Jarlskog factor \cite{Georgi:1979df}) of -3 in the
(2,2)-entry of $Y_\mathrm{d}$ is introduced.
The Yukawa matrix
\begin{eqnarray}\label{Eq:YeTextureExample}
Y_\mathrm{e}=
\begin{pmatrix}
0        & b\epsb^3 & c\epsb^4 \\
b\epsb^3  & -3 \epsb^2 & a\epsb^2 \\
c\epsb^4  & a\epsb^2 & 1 \\
\end{pmatrix} y_\tau
\end{eqnarray}
leads to correct mass ratios for the charged leptons at low energy. Similar
textures have been discussed intensively in the literature and have been used
for the construction of successful fermion mass models in the framework of
unified theories. Here we see an obstacle for QLC with a
charged lepton Yukawa matrix $Y_\mathrm{e}$ similar to the one given in
Eq.~(\ref{Eq:YeTextureExample}): 
because of the Georgi-Jarlskog factor of -3 in the (2,2)-entry of
$Y_\mathrm{e}$,
$\theta_{12}^\mathrm{e}$ is not equal to the Cabibbo angle,
but appears to be three times smaller.
Recall that in order to achieve QLC,
we would require $\theta_{12}^\mathrm{e}$ to be $\sqrt{2}$ larger than
the Cabibbo angle, whereas we have just found it to be three times smaller!
Putting these arguments together we find
$\theta_{12}^\mathrm{e} \approx \tfrac{1}{3} \theta_{\mathrm{C}}$ which with
Eq.~(\ref{Eq:YeContrTexture1_T12_Example}) gives a total correction of
\begin{eqnarray}
\theta_{12} \;\:\approx \;\:\tfrac{\pi}{4} - \tfrac{1}{\sqrt{2}} \tfrac{1}{3}
\theta_{\mathrm{C}}\; .
\end{eqnarray}
In total, the deviation of the lepton mixing $\theta_{12}$ from maximal is
only $\tfrac{1}{\sqrt{2}} \tfrac{1}{3}
\theta_{\mathrm{C}} \approx 3^\circ$.

These, then, are the challenges that must be overcome, if we are to 
achieve QLC. In fact it is possible to overcome these challenges, and
we will see in Sec.~\ref{Sec:Models} how these problems can be solved and how a
QLC relation $\theta_{12}=\tfrac{\pi}{4} - \theta_{\mathrm{C}}$ can be realized
within the framework of quark-lepton-unified models.
The lesson, however, is that the QLC relation is non-trivial
to achieve, and if it is realized in nature, and is not accidental,
then we potentially stand to learn a great deal about the structure
of the underlying unified theory.

\section{Inverted Hierarchy in See-Saw Models} \label{Sec:Inverted}
As noted in \cite{mohfram}, a good starting point towards
obtaining QLC is a Majorana mass matrix for neutrinos that leads
to inverted hierarchy for neutrinos as well as maximal solar
mixing angle. The mass matrix ought to have the form:
\begin{eqnarray}
m^\nu_{\mathrm{LL}}\approx\begin{pmatrix}
0  & b & c \\
b  & 0 & 0 \\
c  & 0 & 0 \\
\end{pmatrix}m_0
\end{eqnarray}
One way to get this form for the mass matrix is to have an
$L_\mathrm{e}-L_\mu-L_\tau$ symmetry \cite{emutau}. One may however
start with a more general mass matrix, following a
procedure introduced in \cite{King:2000ce,King:2002nf},
which we now briefly review. It has been shown that naturalness suggests a
pseudo-Dirac structure of the mass matrix $M_{\mathrm{RR}}$ of the
right-handed neutrinos.\footnote{Such nearly degenerate right-handed
neutrinos are also interesting with respect to resonant
leptogenesis. Note that the right-handed neutrinos can be re-ordered,
which corresponds to a re-ordering of the columns in
$m_{\mathrm{LR}}$.} 
We begin by writing $M_{\mathrm{RR}}$ and the neutrino Dirac
mass matrix $m_{\mathrm{LR}} = Y_\nu v_\mathrm{u}$ as
\begin{eqnarray}
M_{\mathrm{RR}}\approx\begin{pmatrix}
Y  & 0 & 0 \\
0  & 0 & X \\
0  & X & 0 \\
\end{pmatrix},
\quad
m_{\mathrm{LR}} = \begin{pmatrix}
d  & a' & a \\
e  & b' & b \\
f  & c' & c \\
\end{pmatrix}.
\end{eqnarray}
We then impose the 
condition that the pseudo-Dirac right-handed neutrino pair
dominates in the see-saw mechanism \cite{King:2000ce,King:2002nf}:
\begin{eqnarray}\label{Eq:InvHCond_1}
\frac{(d+e+f)^2}{Y} \ll \mbox{Max}\left(\frac{(a'+b'+c')(a+b+c)}{X}\right).
\end{eqnarray}
In \cite{King:2000ce} it was shown that one of the two possibilities
\begin{eqnarray}\label{Eq:InvHCond_2}
a',b,c\gg a,b',c'\quad \mbox{or} \quad a,b',c'\gg a',b,c
\end{eqnarray}
then leads to a natural class of inverted hierarchy models.
For the example $a',b,c\gg a,b',c'$, leads to an inverted hierarchy with
mixing angles,
\begin{eqnarray}
\tan \theta_{23} \approx \frac{c}{b}\; , \quad \theta_{13}\approx \frac{c'b -
b'c}{a'\sqrt{b^2+c^2}}\; , \quad \tan \theta_{12} \approx 1 \; ,
\end{eqnarray}
(and analogous, with primed quantities interchanged with non-primed ones, for
 $a,b',c'\gg a',b,c$). The neutrino mass matrix is given by
\begin{eqnarray}\label{Eq:mLLinverted}
m^\nu_{\mathrm{LL}}\approx\begin{pmatrix}
0  & b & c \\
b  & 0 & 0 \\
c  & 0 & 0 \\
\end{pmatrix} \frac{a' v^2_\mathrm{u}}{X} \; .
\end{eqnarray}
For instance, $M_{\mathrm{RR}}$ and $m_{\mathrm{LR}}$ of the approximate forms
\begin{eqnarray}\label{Eq:InvHierarchyExample}
M_{\mathrm{RR}} = \begin{pmatrix}
Y & 0 & 0 \\ 0 & 0 & X\\ 0 & X & 0 
\end{pmatrix},
\quad
m_{\mathrm{LR}} = \begin{pmatrix}
0  & a' & 0 \\
0  & 0 & b \\
0  & 0 & c \\
\end{pmatrix},
\end{eqnarray}
would lead to the neutrino mass matrix of Eq.~(\ref{Eq:mLLinverted}). 
Of course, small perturbations have to be added in both cases in order to
generate the required small solar mass splitting
\cite{King:2000ce,King:2002nf}.
Note that, although the resulting effective neutrino mass matrix
respects the symmetry $L_\mathrm{e}-L_\mu-L_\tau$, it is not necessary
to assume this symmetry in the construction of this matrix.
In the framework of Pati-Salam models, we will give a possible choice of 
U(1)-charges which lead to mass matrices similar to 
Eq.~(\ref{Eq:InvHierarchyExample}).

\section{QLC in Models of Quark-Lepton Unification}\label{Sec:Models}
We will assume that the symmetry group of our theory contains the
Pati-Salam gauge group
$
G_{422} = \SU (4)_\mathrm{C}\times \SU (2)_\mathrm{L}\times
\SU (2)_\mathrm{R}
$, 
plus an additional flavour symmetry group $F$, which we shall only
partly specify, but which contains a family symmetry 
$\mathrm{U}(1)_\mathrm{X}$. 
Quarks and leptons are unified in the $ \SU (4)_\mathrm{C}$-quartets
$F_{f}$ and
$F^{\ChargeC}_f$ of
$ \SU (4)_\mathrm{C}$, which are doublets of
$\SU (2)_\mathrm{L}$ and $\SU (2)_\mathrm{R}$,
respectively,
\begin{eqnarray}\label{eq:FermionsInPati-SalamModel}
F_i =
\left( \begin{array}{cccc}
u_i & u_i & u_i& \nu_i \\
d_i & d_i & d_i& e_i
\end{array} \right)
, \quad
F_j^\ChargeC =
\left( \begin{array}{cccc}
u_j^\ChargeC & u_j^\ChargeC & u_j^\ChargeC & \nu_j^\ChargeC \\
d_j^\ChargeC & d_j^\ChargeC & d_j^\ChargeC & e_j^\ChargeC
\end{array} \right) .
\end{eqnarray}
$i$ and $j$ are family indices.
The field content of the types of models we are considering in this
section is summarized in Tab.~\ref{Tab:PSfields}.
$\Fla$ and $\Flb$ are flavon fields which we will use in Sec.~\ref{Sec:ModelA}.
They are $G_{422}$-singlets but may be 
charged under the family symmetry $\mathrm{U}(1)_\mathrm{X} \subset F$,
which we will introduce below and which will lead to inverted hierarchy in the
neutrino sector.  

\begin{table}[h]
\begin{center}
\begin{tabular}{|cccccccccccc|}\hline
$\vphantom{\sqrt{\big|}}$field &
$F_1$ & $F_2$ & $F_3$ &
$F_1^\ChargeC$ & $F_2^\ChargeC$ & $F_3^\ChargeC$ &
$h$ & $H$ & $\bar H$ & $\Fla$ & $\Flb$  \\
    \hline
$\vphantom{\sqrt{\big|}^C} \mathrm{SU}(4)_\mathrm{C}$ &
$\boldsymbol{4}$ & $\boldsymbol{4}$ &$\boldsymbol{4}$ &
$\overline{\boldsymbol{4}}$&$\overline{\boldsymbol{4}}$&
$\overline{\boldsymbol{4}}$&
$\boldsymbol{1}$ & $\boldsymbol{4}$ & $\overline{\boldsymbol{4}}$ &
$\boldsymbol{1}$ & $\boldsymbol{1}$ \\
$\vphantom{\sqrt{\big|}}  \mathrm{SU}(2)_\mathrm{L}$&
$\boldsymbol{2}$ & $\boldsymbol{2}$ &$\boldsymbol{2}$ &
$\boldsymbol{1}$ & $\boldsymbol{1}$ &$\boldsymbol{1}$ &
$\boldsymbol{2}$ & $\boldsymbol{1}$ & $\boldsymbol{1}$ & $\boldsymbol{1}$
& $\boldsymbol{1}$ \\
$\vphantom{\sqrt{\big|}}  \mathrm{SU}(2)_\mathrm{R}$&
$\boldsymbol{1}$ & $\boldsymbol{1}$ &$\boldsymbol{1}$ &
$\overline{\boldsymbol{2}}$ &
$\overline{\boldsymbol{2}}$&$\overline{\boldsymbol{2}}$ &
$\boldsymbol{2}$ & $\boldsymbol{2}$ & $\overline{\boldsymbol{2}}$ &
$\boldsymbol{1}$ & $\boldsymbol{1}$
\\
\hline
\end{tabular}
\end{center}
\caption{\label{Tab:PSfields}
Field content assumed in this section.
}
\end{table}

\noindent The Yukawa matrices and the mass matrix $M_\mathrm{RR}$ for the
right-handed
neutrinos will receive contributions from operators of the form
\cite{Allanach:1996hz,King:1998nv}:
\begin{subequations}\begin{eqnarray}
(Y_\mathrm{f})_{ij}&:&
F_i \:F^\ChargeC_j  \:h  \label{Eq:PSYukawaOperators}
\left(\frac{H \bar H}{M_V^2}\right)^n \!\!
\left(\frac{\Fla}{M_V}\right)^{p_{ij}} \!\!
\left(\frac{\Flb}{M_V}\right)^{\bar p_{ij}} , \\
(M_\mathrm{RR})_{ij}&:&
F^\ChargeC_i \:F^\ChargeC_j\:
\frac{H H}{\Lambda}
\left(\frac{H \bar H}{M_V^{2}}\right)^m \!\!
\left(\frac{\Fla }{M_V}\right)^{q_{ij}}\!\!
\left(\frac{\Flb }{M_V} \right)^{\bar q_{ij}} ,
\end{eqnarray}\end{subequations}
where all indices apart from family indices have been suppressed. If the operators
contain $H \bar H$ to some power, various
contractions of the indices are possible. After the Higgs fields $H$, $\bar H$
and $h$ acquire
their vevs, which breaks $G_{422}$ to the SM gauge group $G_{321}$ and
$G_{321}$ to SU(3)$_{C}\times$U(1)$_{\mathrm{el}}$, this can lead to different contributions to the Yukawa couplings of
 up-type quarks, down-type quarks, charged leptons and neutrinos,
 \begin{eqnarray}
 a_{ij}
\left[
  (x_\mathrm{u})_{ij} u_i u_j^\ChargeC h^0_\mathrm{u}
+ (x_\mathrm{d})_{ij} d_i d_j^\ChargeC h^0_\mathrm{d}
+ (x_\mathrm{e})_{ij} e_i e_j^\ChargeC h^0_\mathrm{d}
+ (x_\nu)_{ij}        \nu_i \nu_j^\ChargeC h^0_\mathrm{u}
\right]
\, \delta^m \,\eps^{p_{ij} + \bar p_{ij}}\; .
 \end{eqnarray}
$(x_\mathrm{u})_{ij}$, $(x_\mathrm{d})_{ij}$, $(x_\mathrm{e})_{ij}$ and
$(x_\nu)_{ij}$ are the Clebsch factors for a given operator ${\cal O}$.
Some operators, which will be used in the remainder,
are listed in Tab.~\ref{Tab:SomeOperators} and the corresponding  
Clebsch factors are given in 
Tab.~\ref{Tab:Clebsches} for convenience. An extensive
list of the
operators and the possible Clebsch factors can be found, e.g.,
in \cite{Allanach:1996hz}. We will give an explicit example in 
Eq.~(\ref{Eq:QLCoperators}). 
$a_{ij}$ is an $\mathscr{O}(1)$-factor which arises from generating the
effective
operator. $h^0_\mathrm{u}$ and $h^0_\mathrm{d}$ are the electrically neutral
components of the bi-doublet Higgs $h$.
 Of course, operators can be forbidden, e.g.\ by the horizontal symmetry $F$ 
 or if there is no appropriate
massive field in the full theory for generating this operator.

G$_{422}$ is broken to the SM via vevs $\<H\>=\<\bar{H}\>$ 
of the quartet-Higgses $H$ and $\bar{H}$,
\begin{eqnarray}
H =
\left( \begin{array}{cccc}
0&0&0 & v_H \\
0&0&0&0
\end{array} \right)
, \quad
\bar{H} =
\left( \begin{array}{cccc}
0&0&0 & v_H \\
0&0&0&0
\end{array} \right) .
\end{eqnarray} 
In the following, we will assume a single expansion parameter $\lambda$, where
\begin{eqnarray} \label{Eq:DefDeltaEps}
\eps := \frac{\< \Fla \>}{M_V} = \frac{\< \Flb \>}{M_V} = \lambda\; , \quad
\delta := \frac{\< H \> \<\bar{H}\>}{M_V^2}= \lambda \;,
\end{eqnarray}
with $\lambda \approx 0.22$ being the Cabibbo angle.

\subsection{A Minimal set of Operators that Leads to QLC }\label{Sec:ModelA}

We now present a minimal set of operators, which
will lead to QLC while remaining in agreement with other fermion masses and
mixings. Ideally, one would like to have the choice of operators
dictated entirely by some symmetry group $F$, which will then provide more
fundamental insight into the nature of high scale physics. Our
goal in this note is more modest and we remain content with
simply writing down the minimal set of operators that fulfill our
requirements within an 
$\SU (4)_\mathrm{C}\times \SU (2)_\mathrm{L}\times
\SU (2)_\mathrm{R} \times \U (1)_\mathrm{X}$ 
model, where $\U(1)_\mathrm{X} \subset F$. 
The $\U(1)_\mathrm{X}$ family symmetry will lead to inverted hierarchy 
in the neutrino sector and, with charge assignments\footnote{Note 
that in Pati-Salam models any set of U(1)-charges can be transformed 
to an anomaly-free set of charges leading to the same model.
Thus, anomaly cancellation for the U(1)-symmetry does not restrict
model building in the Pati-Salam framework.} as in 
Tab.~\ref{Tab:U(1)ModelA}, it will 
serve as a starting point for the construction
of the set of operators.
For the set of operators presented in this subsection, 
maximal lepton mixing $\theta_{12}$
as well as large lepton mixing $\theta_{23}$ shall originate from the neutrino 
sector. An alternative route for model building with QLC will be discussed 
 in Appendix~\ref{Sec:ModelB}. 

\begin{table}
\begin{center}
\begin{tabular}{|cccccccccccc|}\hline
$\vphantom{\sqrt{\big|}}$field &
$F_1$ & $F_2$ & $F_3$ & 
$F_1^\ChargeC$ & $F_2^\ChargeC$ & $F_3^\ChargeC$ &
$h$ & $H$ & $\bar H$ & $\Fla$ & $\Flb$  \\
	\hline
$\vphantom{\sqrt{\big|}}  \mathrm{U}(1)_\mathrm{X}$&
-2 &  1 &  1 &
 0 &  1 & -1 &
 0 &  0 &  0 & 1 & -1 \\ \hline
\end{tabular}	
\end{center}
\caption{\label{Tab:U(1)ModelA}
U(1)$_\mathrm{X}$-charges 
(anomaly-free after GS anomaly cancellation)
which help to realize the desired inverted hierarchy in the neutrino sector. It
leads to $M_{\mathrm{RR}}$ and $Y_\nu$ consistent with the requirements of  
Sec.~\ref{Sec:Inverted}. 
However, 
as explained in the text, U(1)$_\mathrm{X}$ should be understood as 
being part of a larger set of family symmetries $F$. 
}
\end{table}

The strategy is the following: 
The choice  of the U(1)$_\mathrm{X}$-charges determines the basic common 
structure of the Yukawa matrices, 
i.e.\ the suppression of the entries by powers of $\eps$ (cf.\ Eq.~(\ref{Eq:DefDeltaEps})). 
Obviously, this structure has to be modified in order to be consistent with 
the charged lepton and quark data at low energy. 
This can be done by forbidding some of the leading 
operators compatible with U(1)$_\mathrm{X}$ symmetry, 
yielding additional suppression of these entries of the Yukawa matrices  
by factors of $\delta$. Where entries of the Yukawa
matrices are allowed at the same order in $\eps$ and in $\delta$, the 
freedom of choosing the operator which gives these entries will be used in
some cases to introduce Clebsch factors, as given in 
Tab.~\ref{Tab:Clebsches}. Using these tools, we arrive at a model with
$Y_\mathrm{u}$, $Y_\mathrm{d}$, $Y_\mathrm{e}$, $Y_\nu$ and 
$M_{\mathrm{RR}}$ given in Tab.~\ref{Tab:ModelA} with the operators defined in
Tab.~\ref{Tab:SomeOperators}. 
We will now discuss how it solves the obstacles for QLC found 
in Sec.~\ref{Sec:TextureExample} and investigate its additional predictions.

\begin{table}
\begin{center}
\begin{tabular}{|ccc|}\hline & & \\[-0.1cm]
$
 Y_{\mathrm{u}}   \!\!\! \:\;=\:\;  \!\! \!\!
\begin{pmatrix}
a_{11}{\cal O}^A\eps^2 \delta^6    &
c_{12}{\cal O}^A  \eps \delta^7 &
b_{13}{\cal O}^A  \eps^3 \delta^6   \\
c_{21}{\cal O}^A\eps \delta^7    &
b_{22}{\cal O}^V \eps^2 \delta^2  &
c_{23}{\cal O}^A\delta^5   \\
b_{31}{\cal O}^A\eps \delta^7    &
b_{32}{\cal O}^V\eps^2 \delta^2    &
a_{33}   \\
\end{pmatrix}
$
& $\rightarrow$ &
$
\begin{pmatrix}
a_{11}\lambda^8  &         c_{12}\lambda^8  & b_{13}\lambda^9 \\
c_{21}\lambda^8  & \sqrt{2}b_{22}\lambda^4  & c_{23}\lambda^5\\
b_{31}\lambda^8  & \sqrt{2}b_{32}\lambda^4  & a_{33} \\
\end{pmatrix}
$
\\[0.8cm] \hline & & \\[-0.4cm]
$
Y_{\mathrm{d}} \!\!\!  \:\;=\:\;  \!\!\!\!
\begin{pmatrix}
a_{11}{\cal O}^A\eps^2 \delta^6    &
b_{12}{\cal O}^W  \eps \delta^3 &
a_{13}{\cal O}^M  \eps^3 \delta   \\
a_{21}{\cal O}^E\eps \delta^3    &
a_{22}{\cal O}^G \eps^2 \delta  &
b_{23}{\cal O}^M \delta^2   \\
b_{31}{\cal O}^A\eps \delta^7    &
a_{32}{\cal O}^M\eps^2 \delta    &
a_{33}   \\
\end{pmatrix}
$
& $\rightarrow$ &
$
\begin{pmatrix}
  a_{11}\lambda^8  & \sqrt{\frac{2}{5}}b_{12}\lambda^4  & \sqrt{2}a_{13}\lambda^4 \\
2 a_{21}\lambda^4  & \frac{2}{\sqrt{5}}a_{22}\lambda^3  & \sqrt{2}b_{23}\lambda^2\\
  b_{31}\lambda^8  &           \sqrt{2}a_{32}\lambda^3   & a_{33} \\
\end{pmatrix}
$
\\[0.8cm] \hline  & & \\[-0.4cm]
$
Y_{\mathrm{e}} \!\!\!  \:\;=\:\;  \!\!\!\!
\begin{pmatrix}
a_{11}{\cal O}^A\eps^2 \delta^6    &
b_{12}{\cal O}^W  \eps \delta^3 &
a_{13}{\cal O}^M  \eps^3 \delta   \\
b_{21}{\cal O}^g\eps \delta^4    &
a_{22}{\cal O}^G \eps^2 \delta  &
b_{23}{\cal O}^M \delta^2   \\
a_{31}{\cal O}^g\eps \delta^4   &
a_{32}{\cal O}^M\eps^2 \delta    &
a_{33}   \\
\end{pmatrix}
$
& $\rightarrow$ &
$
\begin{pmatrix}
        a_{11}\lambda^8  &\!\!\!\!\!\!\!\!\!-3\sqrt{\frac{2}{5}}b_{12}\lambda^4 \!\!\!\!\!\!\! & \sqrt{2}a_{13}\lambda^4 \\
\sqrt{2}b_{21}\lambda^5  & 2\frac{2}{\sqrt{5}}a_{22}\lambda^3  & \sqrt{2}b_{23}\lambda^2\\
\sqrt{2}a_{31}\lambda^5  &            \sqrt{2}a_{32}\lambda^3  & a_{33} \\
\end{pmatrix}
$
\\[0.8cm] \hline  & & \\[-0.4cm]
$
Y_\nu \!\!\!  \:\;=\:\;  \!\!\!\!
\begin{pmatrix}
a_{11}{\cal O}^A\eps^2 \delta^6    &
a_{12}{\cal O}^I  \eps \delta &
b_{13}{\cal O}^A  \eps^3 \delta^6   \\
b_{21}{\cal O}^g\eps \delta^4    &
b_{22}{\cal O}^V \eps^2 \delta^2  &
a_{23}{\cal O}^I \delta   \\
a_{31}{\cal O}^g\eps \delta^4   &
b_{32}{\cal O}^V\eps^2 \delta^2    &
a_{33}    \\
\end{pmatrix}
$
& $\rightarrow$ &
$
\begin{pmatrix}
        a_{11}\lambda^8  &       2 a_{12}\lambda^2    & b_{13}\lambda^9 \\
\sqrt{2}b_{21}\lambda^5  & \sqrt{2}b_{22}\lambda^4  & 2 a_{23} \lambda\\
\sqrt{2}a_{31}\lambda^5  & \sqrt{2}b_{32}\lambda^4  &   a_{33} \\
\end{pmatrix}
$
\\[0.8cm] \hline  & & \\[-0.4cm]
$
M_\mathrm{RR}= \begin{pmatrix}
r_{11}  & r_{12}\eps \delta^2    & r_{13}\eps \delta^2 \\
r_{12}\eps \delta^2 & r_{22}\eps^2 \delta^2 & r_{23} \\
r_{13}\eps \delta^2 & r_{23} & r_{33}\eps^2 \delta^2\\
\end{pmatrix}M_\mathrm{R}
$
& $\rightarrow$ &
$
\begin{pmatrix}
r_{11}  & r_{12}\lambda^3   & r_{13}\lambda^3 \\
r_{12}\lambda^3  & r_{22}\lambda^4 &  r_{23}\\
r_{13}\lambda^3  & r_{23}  & r_{33}\lambda^4 \\
\end{pmatrix}M_\mathrm{R}
$
 \\[0.8cm] \hline
\end{tabular}
\end{center}
\caption{\label{Tab:ModelA}
Complete minimal set of operators for $Y_\mathrm{u}$, $Y_\mathrm{d}$, $Y_\mathrm{e}$, $Y_\nu$ and
$M_{RR}$ that lead to QLC as discussed in Sec.~\ref{Sec:ModelA}. 
${\cal O}$ specifies the type of
operator which yields this contribution, using the convention given in Appendix
1 of \cite{Allanach:1996hz}. The form of the used operators is 
given explicitly in Tab.~\ref{Tab:SomeOperators}. 
The powers of $\delta$ and $\eps$
indicate the order in $H \bar{H}/M_V^2$ and $\Fla/M_V$ or $\Flb/M_V$,
respectively. The resulting Clebsch factors are summarized in 
Tab.~\ref{Tab:Clebsches}.
}
\end{table}

\begin{table}
\begin{center}
\begin{tabular}{|c|c|}\hline
$\vphantom{\sqrt{\big|}}$abbreviation in Tab.~\ref{Tab:ModelA}&  
operator
\\ \hline
$\vphantom{\sqrt{\big|}}$
$
{\cal O}^A \eps^{p_{ij}+\bar{p}_{ij}} \delta^{n+1}
$
&
$
 \frac{1}{M_V^{ p_{ij} + \bar{p}_{ij} + 2(n+1) }}
 (F^a_{i} h_a^y  F^\ChargeC_{j y})_{1}
 \,\Fla^{p_{ij}}\, \Flb^{\bar{p}_{ij}} \,
 \left[(H^z\bar{H}_z)_1\right]^{n+1}
$
\\
$\vphantom{\sqrt{\big|}}$
$
{\cal O}^W \eps^{p_{ij}+\bar{p}_{ij}} \delta^{n+1}
$
&
$
 \frac{1}{M_V^{ p_{ij} + \bar{p}_{ij} + 2(n+1) }}
 (F^a_{i}  F^\ChargeC_{j x})_{15}\, h_a^y \epstensor_{yw} \epstensor^{xz}
 \,(H^w \bar{H}_z)_{15} 
 \,\Fla^{p_{ij}}\, \Flb^{\bar{p}_{ij}} \,
 \left[(H^z\bar{H}_z)_1\right]^n
$
\\
$\vphantom{\sqrt{\big|}}$
$
{\cal O}^G \eps^{p_{ij}+\bar{p}_{ij}} \delta^{n+1}
$
&
$
 \frac{1}{M_V^{ p_{ij} + \bar{p}_{ij} + 2(n+1) }}(F^a_{j}  H^w)_{10}\, h_a^y \epstensor_{yw} \epstensor^{xz}
 \,(F^\ChargeC_{j x} \bar{H}_z)_{10} 
 \,\Fla^{p_{ij}}\, \Flb^{\bar{p}_{ij}} \,
 \left[(H^z\bar{H}_z)_1\right]^n
$
\\
$\vphantom{\sqrt{\big|}}$
$
{\cal O}^g \eps^{p_{ij}+\bar{p}_{ij}} \delta^{n+1}
$
&
$
 \frac{1}{M_V^{ p_{ij} + \bar{p}_{ij} + 2(n+1) }}
 (F^a_{i}  \bar{H}_z)_{1}\, h_a^y 
 \,(F^\ChargeC_{j y} H^z)_{1}
\,\Fla^{p_{ij}}\, \Flb^{\bar{p}_{ij}} \,
 \left[(H^z\bar{H}_z)_1\right]^n
$ 
\\
$\vphantom{\sqrt{\big|}}$
$
{\cal O}^V \eps^{p_{ij}+\bar{p}_{ij}} \delta^{n+1}
$
&
$
 \frac{1}{M_V^{ p_{ij} + \bar{p}_{ij} + 2(n+1) }}(F^a_{i}  F^\ChargeC_{j w})_{1}\, h_a^y 
 \,(H^w \bar{H}_y)_{1}
\,\Fla^{p_{ij}}\, \Flb^{\bar{p}_{ij}} \,
 \left[(H^z\bar{H}_z)_1\right]^n
$ 
\\
$\vphantom{\sqrt{\big|}}$
$
{\cal O}^M \eps^{p_{ij}+\bar{p}_{ij}} \delta^{n+1}
$
&
$
 \frac{1}{M_V^{ p_{ij} + \bar{p}_{ij} + 2(n+1) }}
(F^a_{i}  F^\ChargeC_{j x})_{1}\, h_a^y \epstensor_{yw} \epstensor^{xz}
 \,(H^w \bar{H}_z)_{1}
\,\Fla^{p_{ij}}\, \Flb^{\bar{p}_{ij}} \,
 \left[(H^z\bar{H}_z)_1\right]^n
$
\\
$\vphantom{\sqrt{\big|}}$
$
{\cal O}^E \eps^{p_{ij}+\bar{p}_{ij}} \delta^{n+1}
$
&
$
 \frac{1}{M_V^{ p_{ij} + \bar{p}_{ij} + 2(n+1) }}(F^a_{j}  H^w)_{6}\, h_a^y \epstensor_{yw} \epstensor^{xz}
 \,(F^\ChargeC_{j x} \bar{H}_z)_{6} 
 \,\Fla^{p_{ij}}\, \Flb^{\bar{p}_{ij}} \,
 \left[(H^z\bar{H}_z)_1\right]^n
$
\\
$\vphantom{\sqrt{\big|}}$
$
{\cal O}^I \eps^{p_{ij}+\bar{p}_{ij}} \delta^{n+1}
$
&
$
 \frac{1}{M_V^{ p_{ij} + \bar{p}_{ij} + 2(n+1) }}
(F^a_{i}  \bar{H}_y)_{1}\, h_a^y 
 \,(F^\ChargeC_{j w} H^w)_{1}
\,\Fla^{p_{ij}}\, \Flb^{\bar{p}_{ij}} \,
 \left[(H^z\bar{H}_z)_1\right]^n
$
\\
\hline
\end{tabular}
\end{center}
\caption{\label{Tab:SomeOperators}
Explicit form of the operators used within the set of operators which
leads to QLC (Tab.~\ref{Tab:ModelA}). Only 
indices of SU(2)$_\mathrm{L}$ and 
SU(2)$_\mathrm{R}$ and family indices $i,j$ are shown 
explicitly and brackets $(X Y)_R$ indicates the product of the fields such that 
the expression transforms in the representation R of
SU(4)$_\mathrm{C}$. The complete set of possible index contractions for 
the operators of Eq.~(\ref{Eq:PSYukawaOperators}) can be found, e.g., in Appendix 1 of \cite{Allanach:1996hz}.
After symmetry breaking of $G_{422}$ to the SM, the operators 
lead to Clebsch factors given in 
Tab.~\ref{Tab:Clebsches}.  
}
\end{table}

\begin{table}
\begin{center}
\begin{tabular}{|l|ccccc|}\hline
${\cal O}$ $\vphantom{\sqrt{\big|}}$& $x_\mathrm{u}$ & $x_\mathrm{d}$ &
$x_\mathrm{e}$ & $x_\nu$ & comment \\ \hline
${\cal O}^A$ $\vphantom{\sqrt{\big|}}$& 1 & 1 & 1 & 1
& $x_\mathrm{u}=x_\mathrm{d}=x_\mathrm{e}=x_\nu$ \\
%
%
${\cal O}^W$ $\vphantom{\sqrt{\big|}}$& 0 & $\sqrt{\frac{2}{5}}$ & $-3
\sqrt{\frac{2}{5}}$ & 0
& $x_\mathrm{e}/x_\mathrm{d} = -3$\\
${\cal O}^G$ $\vphantom{\sqrt{\big|}}$& 0 & $\frac{2}{\sqrt{5}}$ &
$\frac{4}{\sqrt{5}}$ & 0
& $x_\mathrm{e}/x_\mathrm{d} = 2$\\
${\cal O}^g$ $\vphantom{\sqrt{\big|}}$& 0 & 0 & $\sqrt{2}$ & $\sqrt{2}$
& only $x_\mathrm{e}=x_\nu \not=0$ \\
${\cal O}^V$ $\vphantom{\sqrt{\big|}}$& $\sqrt{2}$ & 0 & 0 & $\sqrt{2}$
& only $x_\mathrm{u}=x_\nu \not=0$ \\
${\cal O}^M$ $\vphantom{\sqrt{\big|}}$& 0 & $\sqrt{2}$ & $\sqrt{2}$ & 0
& only $x_\mathrm{d}=x_\mathrm{e} \not=0$ \\
${\cal O}^E$ $\vphantom{\sqrt{\big|}}$& 0 & 2 & 0 & 0
& only $x_\mathrm{d} \not=0$\\
%
%
${\cal O}^I$ $\vphantom{\sqrt{\big|}}$& 0 & 0 & 0 & 2
& only $x_\nu \not=0$\\\hline
\end{tabular}
\end{center}
\caption{\label{Tab:Clebsches}
Clebsch factors $x_i$
for the operators defined in Tab.~\ref{Tab:SomeOperators}, normalized to 
$\sum_i x_i^2 = 4$.
A complete list of operators and Clebsch factors can be found
in the appendix of \cite{Allanach:1996hz}.
}
\end{table}

\subsubsection{Realizing QLC}
Before we discuss the predictions in detail, let us point out the main 
properties of the set of operators given in Tab.~\ref{Tab:ModelA}, 
which allow to realize QLC in model of quark-lepton unification. 
Omitting common $\mathscr{O}(1)$-coefficients and
treating tiny elements as zero, $Y_\mathrm{e}$, 
$Y_\mathrm{d}$ and $m_\mathrm{LL}$ are given by (cf.\ Tab.~\ref{Tab:ModelA})
\begin{eqnarray}\label{Eq:ModelAMainFeatures}
\begin{array}{lll}
\!\!\!\!\!   Y_{\mathrm{d}}\sim
\begin{pmatrix} 
  0  & 1 \lambda^4  &\lambda^4 \\ 
*  & 1 \lambda^3  & \lambda^2\\
*  &           *   & 1 \\
\end{pmatrix} 
\;,
&
Y_\mathrm{e}\sim
\begin{pmatrix} 
        0  &-3\lambda^4 & \lambda^4 \\ 
*  & 2\lambda^3  &\lambda^2\\
*  &         *  & 1 \\
\end{pmatrix}\;,
&
m_\mathrm{LL} \sim \begin{pmatrix} 
0  & m    &  m' \\ 
m  & 0 &0  \\
m'  &0 &0 \\
\end{pmatrix},
 \\
\end{array}
\end{eqnarray}
where we remind the reader that $\lambda \approx \theta_C$.
For QLC, the relation between $Y_\mathrm{e}$ and $Y_\mathrm{d}$ is
crucial. 
Here we have chosen the following operators for the (1,2)- and (2,2)-entries 
in $Y_\mathrm{e}$ and $Y_\mathrm{d}$:
\begin{eqnarray}\label{Eq:QLCoperators}
(Y_{\mathrm{e},\mathrm{d}})_{12} &:& b_{12} {\cal O}^W \eps \delta^3 
\quad \leftrightarrow \quad
 \frac{b_{12}}{M_V^7}(F^a_{1}  F^\ChargeC_{2 x})_{15}\, h_a^y \epstensor_{yw} \epstensor^{xz}
 \,(H^w \bar{H}_z)_{15} \,\left[(H^z\bar{H}_z)_1\right]^2 \Fla \; ,
\nonumber \\
(Y_{\mathrm{e},\mathrm{d}})_{22} &:& a_{22} {\cal O}^G \eps^2 \delta 
\quad \leftrightarrow \quad 
\frac{a_{22}}{M_V^4}(F^a_{2}  H^w)_{10}\, h_a^y \epstensor_{yw} \epstensor^{xz}
 \,(F^\ChargeC_{2 x} \bar{H}_z)_{10} \,\Flb^2 \; ,
\end{eqnarray}
where only indices of SU(2)$_\mathrm{L}$ and 
SU(2)$_\mathrm{R}$ are shown 
explicitly and where brackets $(X Y)_R$ indicates the product of the fields such that 
the expression transforms in the representation R of
SU(4)$_\mathrm{C}$.\footnote{The product is normalized such that the resulting Clebsch factors are in the convention 
used in Tab.~\ref{Tab:Clebsches}, i.e.\ $\sum_i x_i^2 = 4$. More details can be
found, e.g.\, in Appendix 1 of \cite{Allanach:1996hz}.} 
After symmetry breaking of $G_{422}$ to the SM, the operators 
lead to Clebsch factors $-3$ for 
$(Y_\mathrm{e})_{12}/(Y_\mathrm{d})_{12}$ and 
$2$ for $(Y_\mathrm{e})_{22}/(Y_\mathrm{d})_{22}$ (cf.\ 
Tab.~\ref{Tab:Clebsches}).  
The above 
choice of Clebsch factors is a crucial feature of the model with respect
to QLC. It solves the problems found in Sec.~\ref{Sec:TextureExample},
and in particular leads to a charged lepton mixing angle 
approximately $\sqrt{2}$ larger than the Cabibbo angle:
$
\theta_{12}^\mathrm{e} = \tfrac{3}{2} \theta_\mathrm{C} \approx \sqrt{2} 
\theta_\mathrm{C}.
$
Using Eq.~(\ref{Eq:YeContrTexture1_T12_Example}), this 
leads to the desired relation $\theta_{12} \approx \tfrac{\pi}{4} -
\theta_\mathrm{C}$ at high energy $M_\mathrm{X}$.\footnote{From Tab.~\ref{Tab:ModelA} 
we see that the contribution from $Y_\mathrm{u}$ to the Cabibbo angle is
very small.}  

We also see immediately that 
the Clebsch factor $2$ for $(Y_\mathrm{e})_{22}/(Y_\mathrm{d})_{22}$  
leads to the prediction $m_\mu/m_s \approx 2$ at $M_{\mathrm{X}}$. 
Furthermore, from Eq.~(\ref{Eq:YeContrTexture1_T13_Example}), we see that an additional
prediction is $\theta_{13} \approx \theta_\mathrm{C}$ at $M_\mathrm{X}$. 
Contrary to these main features, 
some details of the model such as the choice of some of the other operators 
are quite arbitrary.

\subsubsection{Predictions at $\boldsymbol{M_{\mathrm{X}}}$}
The model given in Tab.~\ref{Tab:ModelA}  
can accommodate the estimated parameters of the quark and 
charged sector at the unification scale $M_\mathrm{X}=10^{16}$ GeV (e.g.\ from \cite{King:1998nv}) by 
choosing appropriate values for the $\mathscr{O}(1)$-coefficients. This is
demonstrated analytically in Tab.~\ref{Tab:O(1)CoeffAndQuarks}. In addition, 
we obtain a prediction for $m_s$, which we will discuss separately below.

\begin{table}
\begin{center}
\begin{tabular}{|c|c|c|c|}\hline
$\vphantom{\sqrt{\big|}}$ quantity & at $M_X$ given by & required at $M_X$ \cite{King:1998nv}& determined at $M_X$
\\ \hline
$y_\mathrm{t}=y_\mathrm{b}=y_\tau$ & $\approx a_{33}$ & $\approx 0.68$ &
$a_{33}\approx 0.68$ \\
$\theta^\mathrm{CKM}_{23} \approx V_{cb}$ & $\approx\frac{\sqrt{2}b_{23}}{a_{33}} \lambda^2$ & $\approx 0.032$ &
$b_{23}\approx 0.32$ \\
$\theta^\mathrm{CKM}_{13} \approx V_{ub}$ & $\approx\frac{\sqrt{2}a_{13}}{a_{33}} \lambda^4$ &
$\approx 0.0028$ &  $a_{13}\approx 0.57$ \\
$y_\mu$ & $\approx2\frac{2}{\sqrt{5}} a_{22}\lambda^3 $ & $\approx 3.2 \cdot 10^{-2}$ &
 $ a_{22} \approx 1.7$ \\
$\theta^\mathrm{CKM}_{12} = \theta_\mathrm{C} \approx V_{us}$ & $\approx\frac{b_{12}}{\sqrt{2} a_{22}} \lambda $ & $\lambda$ & $b_{12}\approx 2.4$
 \\
$y_\mathrm{c}$ & $\approx\sqrt{2} b_{22}\lambda^4$ & $\approx 1.5 \cdot 10^{-3}$ & $b_{22}\approx 0.45$
\\
$y_\mathrm{u}$ & $\approx a_{11}\lambda^8 $ & $\approx 4.7 \cdot 10^{-6}$ &
$a_{11} \approx 0.86 $
\\
$y_\mathrm{d}$ &  $\approx\frac{\sqrt{2} a_{21} b_{12}}{a_{22}} \lambda^5$ &
$\approx 3.2 \cdot 10^{-4}$ & $ a_{21} \approx 0.31$
\\
$y_\mathrm{e}$ & $\approx\frac{3 b_{21} b_{12}}{2 a_{22}} \lambda^6$ &
$\approx 1.5 \cdot 10^{-4}$ & $b_{21} \approx 0.62$  \\
 \hline
\end{tabular}
\end{center}
\caption{\label{Tab:O(1)CoeffAndQuarks}
Accommodating the estimated quark data
and the charged lepton masses at
$M_\mathrm{X}\approx 10^{16}$ GeV (from \cite{King:1998nv}, for large $\tan \beta \approx 40$)
by determining the $\mathscr{O}(1)$-coefficients for the
model defined in Tab.~\ref{Tab:ModelA}.
In addition, the model predicts $m_\mu/m_s \approx 2$ at $M_{\mathrm{X}}$,
as discussed in the text -- and of course the desired QLC relation
$\theta_{12} \approx \tfrac{\pi}{4} - \theta_\mathrm{C}$
and additionally
$\theta_{13}=\lambda_\mathrm{C}$ at $M_{\mathrm{X}}$. We have ignored the quark
sector CP phase $\delta$.
}
\end{table}

Let us now consider the lepton sector.
In the neutrino sector, $Y_{\nu}$ and $M_{RR}$ in Tab.~\ref{Tab:ModelA}
have the approximate forms of the matrices in Eq.~(\ref{Eq:InvHierarchyExample}),
with the condition that the pseudo-Dirac right-handed neutrino
pair dominates, and so predicts the required effective neutrino
mass matrix in Eq.~(\ref{Eq:mLLinverted}). This leads to an inverted
neutrino mass hierarchy with almost maximal $\theta_{12}^\nu \approx \pi/4$ 
due to the
approximate Pseudo-Dirac structure in the 1-2 submatrix of
$m_\mathrm{LL}$ in Eq.~(\ref{Eq:mLLinverted}). 
Furthermore, $\theta_{23}^\nu$ is large and $\theta_{13}^\nu$ is generated by 
sub-leading entries in $m_\mathrm{LL}$ (not displayed in 
Eq.~(\ref{Eq:ModelAMainFeatures})) and is thus very small. 
The resulting
mass matrix $m_\mathrm{LL}$ of the light neutrinos is given by
\begin{eqnarray}
m_\mathrm{LL}= \begin{pmatrix}
\mathscr{O}(\lambda^4)     & 4 a_{12} a_{23} \lambda^2 &  2 a_{12} a_{33} \lambda\\
4 a_{12} a_{23}  \lambda^2 & \mathscr{O}(\lambda^3)    &\mathscr{O}(\lambda^3)  \\
2 a_{12} a_{33} \lambda    &\mathscr{O}(\lambda^3)     &\mathscr{O}(\lambda^3) \\
\end{pmatrix}  \frac{\lambda v_\mathrm{u}^2}{r_{23} M_\mathrm{R}} .
\end{eqnarray}
$\theta_{12}^\nu$ is very close to maximal as desired, $\theta_{13}^\nu$ is tiny
and $\arctan \theta_{23}^\nu \approx a_{33}/(2 a_{23} \lambda)$.
Thus, $a_{33} \approx 2 a_{23}\lambda$ and with $a_{33} \approx 0.68$ from
Tab.~\ref{Tab:O(1)CoeffAndQuarks}, $a_{23} \approx 1.5$ is
required for an approximately maximal mixing $\theta_{23}^\nu$.
Furthermore, in order to have $2 a_{12} a_{33}\lambda$ of order $1$,
we choose $a_{12} \approx 2$. All three right-handed neutrinos have masses around
 $M_\mathrm{R}\approx 10^{14}$ for generating the atmospheric mass squared
 difference $\Delta m^2_{31}$ of the right order.
The solar mass squared difference $\Delta m^2_{21}$ is generated from the corrections
to the leading order structure in $m_\mathrm{LL}$.
With $\mathscr{O}(\lambda^3)$
entries in the sub-leading elements of $m_\mathrm{LL}$, it is roughly of the
order $2 \Delta m^2_{31} \lambda^3$ and a value
consistent with experiment is produced for many choices of yet undetermined
$\mathscr{O}(1)$-coefficients.\footnote{Some care has to be taken
when choosing the
$\mathscr{O}(1)$-coefficients that, in the convention with $m_2 > m_1$,
$\theta_{12}$ is $< \tfrac{\pi}{4}$.}
For the corrections from the charged lepton sector, let us consider
$Y_\mathrm{d}$ and $Y_\mathrm{e}$
(Tab.~\ref{Tab:ModelA}, or Eq.~(\ref{Eq:ModelAMainFeatures})).
Diagonalizing $Y_\mathrm{d}$ by
$U_{d_{\mathrm{L}}} Y_\mathrm{d} U^\dagger_{d_{\mathrm{R}}}$
with $U^\dagger_{d_{\mathrm{L}}} = R^q_{23} U^q_{13} R^q_{12} \approx U_\mathrm{CKM}$, we read
off $\theta_{12}^q = \theta_{\mathrm{C}} \approx \lambda$,
$\theta_{13}^q \sim \lambda^4$ and $\theta_{13}^q \sim \lambda^2$.
Diagonalizing $Y_\mathrm{e}$ by
$U_{e_{\mathrm{L}}} Y_\mathrm{e} U^\dagger_{e_{\mathrm{R}}}$
with $U^\dagger_{e_{\mathrm{L}}} = R^\mathrm{e}_{23} U^\mathrm{e}_{13}
R^\mathrm{e}_{12}$, we obtain
\begin{eqnarray}\label{Eq:MixingsThetaE_ModelA}
\theta_{12}^\mathrm{e} \approx \tfrac{3}{2} \lambda \approx \tfrac{3}{2}\theta_{\mathrm{C}} \; , \quad
\theta_{13}^\mathrm{e} \sim \lambda^4 \; , \quad
\theta_{23}^\mathrm{e} \sim \lambda^2 \; .
\end{eqnarray}
Assuming $\theta_{23}^\nu \approx \pi/4$,
$\theta_{13}^\nu \approx 0$ and treating $\theta_{13}^q$ as
$\approx 0$, this gives the high energy predictions
(using Eqs.~(\ref{Eq:YeContributions}) and (\ref{Eq:MixingsThetaE_ModelA}))
\begin{subequations}\label{Eq:YeContrModelA}
\begin{eqnarray}
s_{23} &\approx& s_{23}^\nu - \tfrac{1}{\sqrt{2}}  \theta_{23}^\mathrm{e}
\;\;  \rightarrow \;\;
\theta_{23} \;\:\approx \;\:\theta_{23}^\nu - \theta_{23}^\mathrm{q}
 \; ,\\
\theta_{13} &\approx& \tfrac{1}{\sqrt{2}}  \theta_{12}^\mathrm{e}
\;\;  \rightarrow \;\;
          \theta_{13}   \;\:\approx \;\:  1.06 \,\theta_\mathrm{C}\; ,\\
 s_{12} &\approx& \tfrac{1}{\sqrt{2}}
           - \tfrac{1}{2} \theta_{12}^\mathrm{e}
         \;\;  \rightarrow \;\;
\theta_{12} \;\:\approx \;\:   \tfrac{\pi}{4} - \tfrac{1}{\sqrt{2}}\theta_{12}^\mathrm{e}  
\;\:\approx \;\: \tfrac{\pi}{4} - 1.06 \,\theta_\mathrm{C}
 \; ,
\end{eqnarray}
\end{subequations}
where we have written $\tfrac{3}{2} \frac{1}{\sqrt{2}}=1.06$. 
Therefore the approximate QLC relation
$\theta_{12} \approx \tfrac{\pi}{4} - 1.06 \,\theta_\mathrm{C}$ is realized 
due to the choice of the operators which
give the (1,2)- and (2,2)-entries in $Y_\mathrm{e}$ and
$Y_\mathrm{d}$, yielding Clebsch factors $-3$ for
$(Y_\mathrm{e})_{12}/(Y_\mathrm{d})_{12}$ and
$2$ for $(Y_\mathrm{e})_{22}/(Y_\mathrm{d})_{22}$.

\subsubsection{RG Modification of QLC and other Corrections}
The RG analysis, including the running of the neutrino parameters
with successively integrating out the right-handed neutrinos,
can be performed conveniently
using the software packages \texttt{REAP/MPT} introduced in
\cite{Antusch:2005gp}.\footnote{As can be seen e.g.\ 
from the analytical formulae of \cite{Antusch:2003kp,Antusch:2005gp},
in the model considered here with an inverted hierarchy and 
with a Majorana parity between $m_1$ and
$m_2$, the RG effects zeroth order in $\theta_{13}$ are strongly suppressed and 
corrections proportional to $\theta_{13}$ play the dominant role.
} The RG corrections to the mixing angles in our model
are (assuming $M_{\mathrm{SUSY}}\approx 1$ TeV, $\tan \beta \approx 40$):
\begin{subequations}\begin{eqnarray}
\theta_{12}(M_\mathrm{X}) - \theta_{12}(M_Z) &\approx& 0.8^\circ \; ,\\
\theta_{13}(M_\mathrm{X}) - \theta_{13}(M_Z) &\approx& 0.5^\circ \; ,\\
\theta_{23}(M_\mathrm{X}) - \theta_{23}(M_Z) &\approx& 2.9^\circ \;.
\end{eqnarray} \end{subequations}
In particular, the QLC relation is slightly modified by $\approx 0.8^\circ$ due
to the RG correction. In
addition, there are corrections within the model itself, which can lead to
a deviation from QLC. However, the latter corrections can give deviations
in both directions (+/-), whereas the RG running shifts the prediction to
smaller values. The predictions of the model at
$M_\mathrm{X}$ and at $M_Z$ are summarized in Tab.~\ref{Tab:PredModelA}.

\begin{table}[h]
\begin{center}
\begin{tabular}{|c|c|c|}\hline
$\vphantom{\sqrt{\big|}}$quantity&                     $\theta_{12}$ &
$\theta_{13}$ 
\\ \hline
$\vphantom{\sqrt{\big|}}$prediction at $M_\mathrm{X}$&  $\tfrac{\pi}{4} -1.06\,\theta_\mathrm{C}$ &
$1.06\,\theta_\mathrm{C}$
\\ \hline

$\vphantom{\sqrt{\big|}}$prediction at $M_Z$&
$ \tfrac{\pi}{4} -1.06\,\theta_\mathrm{C} - 0.8^\circ \approx 30.5^\circ$ &
$1.06\,\theta_\mathrm{C} - 0.5^\circ \approx 13.2^\circ$ 
\\
\hline
\end{tabular}
\end{center}
\caption{\label{Tab:PredModelA}
Predictions for the lepton mixing angles at the unification scale $M_\mathrm{X} \approx 10^{16}$ GeV
and at low energy
$M_Z$ for the model discussed in
Sec.~\ref{Sec:ModelA}. RG corrections shift the QLC
prediction to slightly smaller values. The factor $1.06$ is equal to
$1.5/\sqrt{2}$. We have used $\sin (\theta_{\mathrm{C}})= 0.224$. 
In addition to the corrections shown in the table, there are corrections
from the model itself, e.g.\ from sub-leading higher-dimensional operators.
}
\end{table}

\section{Conclusions}
We have studied possible ways to realize the QLC  
relation $\theta_{12}\simeq \pi/4 - \theta_{\mathrm{C}}$
between the Cabibbo angle and the solar
mixing angle in realistic quark-lepton 
unification theories based on the Pati-Salam gauge group 
SU(2)$_\mathrm{L}\times$ SU(2)$_\mathrm{R} \times$ SU(4)$_\mathrm{C}$. 
This represents the  
the first attempt at a unified model at the GUT scale 
capable of predicting the QLC relation.
The Pati-Salam gauge group is necessary to relate the
charged lepton mixing angles to the down-quark mixing angles,
which in turn dominate the contribution to the Cabibbo angle,
allowing the QLC relation to emerge, albeit in a non-trivial way.
If such a QLC relation, which is consistent with 
recent observations, is in fact realized and
if it is not accidental, this would point towards quark-lepton 
unification and could tell us a great deal about the detailed 
structure of the unified theory. 

One necessary ingredient for QLC is maximal mixing 
$\theta^\nu_{12}=\pi/4$ from the neutrino sector.  
We have therefore started from an inverted hierarchy 
structure of the neutrino mass matrix, constructed via the see-saw mechanism, 
which predicts $\theta^\nu_{12}=\pi/4$ and $\theta^\nu_{13}=0$. 
Although the effective neutrino mass matrix has a form which is invariant under
$L_\mathrm{e}-L_\mu-L_\tau$, we have not imposed this symmetry in 
the high energy theory. In fact we have instead relied on a different
$U(1)_\mathrm{X}$ family symmetry as playing a leading role in
achieving the inverted hierarchy structure.

The QLC relation then has to be realized from the contribution to the lepton
mixing matrix of the charged leptons. We have analyzed the generic problems with
realizing QLC in models where the
$SU(4)_C$ quark-lepton unification scale is close to the GUT
scale. In particular, if $\theta^\nu_{12}=\pi/4$ and 
$\theta^\nu_{23}\approx\pi/4$ are generated by the neutrino mass matrix
$m_{\mathrm{LL}}$, $\theta_{12}^\mathrm{e}$ has to be about $\sqrt{2}$ larger
than $\theta_{\mathrm{C}}$ in order to lead to the QLC relation. 
Furthermore, in models of quark-lepton unification, realizing the correct 
ratio of the strange mass to the muon mass at high energy typically requires 
an operator which introduces a group theoretical Clebsch factor 
(Georgi-Jarlskog factor) at the (2,2)-entry of the charge lepton Yukawa matrix. 
Such a Georgi-Jarlskog factor
also affects the relation between $\theta_{12}^\mathrm{e}$ and the 
 Cabibbo angle, making $\theta_{12}^\mathrm{e}$ three times
 larger than $\theta_{\mathrm{C}}$. 
 
In this letter we have identified a minimal set of operators
within the framework of Pati-Salam models which can lead to
approximate QLC while remaining consistent with other known data.
A direct prediction of our approach is that the neutrino mass hierarchy
is of the inverted type. It is instructive to note that in this model
there are significant corrections to exact QLC coming from
a variety of different sources. In order to compensate for the
$1/\sqrt{2}$ suppression referred to above, we have enhanced the charged lepton mixing angle
by a factor of $3/2$, leading to an approximate QLC relation
$\theta_{12}\approx \pi/4 - 1.06\,\theta_C$. In addition RG 
corrections reduce the prediction by a further $0.8^\circ$, resulting in the
final prediction $\theta_{12}\approx 30.5^\circ$, with some theoretical error. 
A generic prediction of our approach is $\theta_{13}\approx \theta_C$,
where in this case the factor of 1.06 enhancement approximately cancels
the RG correction leading to the stated prediction.
These predictions allow these models to be confirmed or excluded
by the current generation of neutrino oscillation experiments.
We have also commented in an Appendix on alternative classes of QLC
models with a lopsided charged lepton mass matrix, which lead
to somewhat different predictions.

\section*{Acknowledgements}
Two of us (S.~A. and S.~F.~K) acknowledge support from PPARC by the grant 
No.\ PPA/G/O/2002/00468 and the work of R.~N.~M. is supported by NSF grant
No.\ PHY-0354401. Two of us (S.~F.~K. and R.~N.~M.) are grateful to the 
organizers of the 2004 Nobel Symposium on Neutrinos at Engkoping, Sweden 
for providing a congenial atmosphere where this work was started.

\section*{Appendix}
\appendix

\renewcommand{\thesection}{\Alph{section}}
\renewcommand{\thesubsection}{\Alph{section}.\arabic{subsection}}
\def\theequation{\Alph{section}.\arabic{equation}}
\renewcommand{\thetable}{\arabic{table}}
\renewcommand{\thefigure}{\arabic{figure}}
\setcounter{section}{0}
\setcounter{equation}{0}

\section{Our Conventions}\label{conventions}
For the mass matrix of the charged leptons $m_{\mathrm{E}}=Y_\mathrm{e}
v_\mathrm{d}$, where
$v_\mathrm{d} = \< h^0_\mathrm{d}\>$,
defined by
$\mathcal{L}_\mathrm{e}=-(m_{\mathrm{E}})_{fg} \overline{ e^f_{\mathrm{L}}} 
e^g_{\mathrm{R}}$ + h.c., 
and for the neutrino mass
matrix $m^\nu_{LL}$, defined by $\mathcal{L}_\nu = 
(m^\nu_{\mathrm{LL}})_{fg} \overline{\nu^f_\mathrm{L}}
\nu_\mathrm{L}^{\ChargeC g}$ + h.c.,
the change from flavour basis to mass
eigenbasis can be performed with the unitary diagonalization matrices
$U_{e_\mathrm{L}},U_{e_\mathrm{R}}$ and
$U_{\nu_\mathrm{L}}$ by
\begin{eqnarray}\label{eq:DiagMe}
U_{e_\mathrm{L}} \, m_{\mathrm{E}} \,U^\dagger_{e_\mathrm{R}} =
\left(\begin{array}{ccc}
\!m_e&0&0\!\\
\!0&m_\mu&0\!\\
\!0&0&m_\tau\!
\end{array}
\right)\! , \quad
U_{\nu_\mathrm{L}} \,m^\nu_{\mathrm{LL}}\,U^T_{\nu_\mathrm{L}} =
\left(\begin{array}{ccc}
\!m_1&0&0\!\\
\!0&m_2&0\!\\
\!0&0&m_3\!
\end{array}
\right)\! .
\end{eqnarray}
The MNS matrix is then given by
\begin{eqnarray}\label{Eq:MNS_Definition}
U_{\mathrm{MNS}} = U_{e_\mathrm{L}} U^\dagger_{\nu_\mathrm{L}}\; .
\end{eqnarray}
We use the parameterization
$
U_{\mathrm{MNS}} = R_{23} U_{13} R_{12} P_0
$
with $R_{23}, U_{13}, R_{12}$ and $P_0$ being defined as
\begin{align}\label{eq:R23U13R12P0}
R_{12}:=
\left(\begin{array}{ccc}
  c_{12} & s_{12} & 0\\
  -s_{12}&c_{12} & 0\\
  0&0&1\end{array}\right)
  , \:&
\quad U_{13}:=\left(\begin{array}{ccc}
   c_{13} & 0 & s_{13}e^{-i\delta}\\
  0&1& 0\\
  - s_{13}e^{i\delta}&0&c_{13}\end{array}\right)  ,  \nonumber \\
R_{23}:=\left(\begin{array}{ccc}
 1 & 0 & 0\\
0&c_{23} & s_{23}\\
0&-s_{23}&c_{23}
 \end{array}\right)
  , \:&
 \quad P_0:=
 \begin{pmatrix}
 1&0&0\\0&e^{i\beta_2}&0\\0&0&e^{i\beta_3}
  \end{pmatrix}
\end{align}
and where $s_{ij}$ and $c_{ij}$ stand for $\sin (\theta_{ij})$ and $\cos
(\theta_{ij})$, respectively.
The matrix $P_0$ contains the possible Majorana
phases $\beta_2$ and $\beta_3$.
 $\delta$ is the Dirac CP phase relevant for neutrino oscillations.

\section{Alternative Approach: Lopsided
$\boldsymbol{Y_\mathrm{e}}$}\label{Sec:ModelB}
Let us now discuss an alternative route to models with QLC, where nearly maximal
lepton mixing $\theta_{23}$ stems from the charged lepton sector. Single maximal
lepton mixing $\theta_{12}$ is then generated by $m_\mathrm{LL}$.
This has the
following advantage for constructing models with QLC:
with $U_{e_\mathrm{L}} \approx
R^\mathrm{e}_{23}(\pi/4) R^\mathrm{e}_{12}(\theta_{\mathrm{C}})$,
the rotations are already in the correct order for the MNS
matrix,\footnote{The very small neutrino mixings $\theta_{23}^\nu$ and $\theta_{13}^\nu$ are
treated as $0$.
With respect to QLC in lopsided models, it is convenient to perform
the 1-2 rotation $R^\mathrm{e}_{12}$ first when diagonalizing $Y_\mathrm{e}$,
followed by the rotations $U^\mathrm{e}_{13}$ and $R^\mathrm{e}_{23}$.
Note that this differs somewhat from the usual convention.}
\begin{eqnarray}
U_\mathrm{MNS} \;\approx\;
R^\mathrm{e}_{23} (\tfrac{\pi}{4}) U^\mathrm{e}_{13}
R^\mathrm{e}_{12}(\theta_\mathrm{C}) R^\nu_{12}(\tfrac{\pi}{4})  P_0^\nu
\;\approx\;
R^\mathrm{e}_{23} (\tfrac{\pi}{4}) U^\mathrm{e}_{13}
R_{12}(\tfrac{\pi}{4} - \theta_\mathrm{C})  P_0^\nu \; .
\end{eqnarray}
In particular,
we do not obtain an unwanted factor of $1/\sqrt{2}$ for the relation between
$\theta_\mathrm{C}$ and
the correction to maximal lepton mixing $\theta_{12}$ (cf.~Eq.~(\ref{Eq:ChargedLContrToT12_Example}).
Omitting common $\mathscr{O}(1)$-coefficients and
treating very small elements as zero, QLC can be achieved with $Y_\mathrm{e}$,
$Y_\mathrm{d}$ and $m_\mathrm{LL}$ of the form
\begin{eqnarray}\label{Eq:ModelBMainFeatures}
\begin{array}{lll}
\!\!\!\!\!   Y_{\mathrm{d}}=
\begin{pmatrix}
  0  & 1 \lambda^4  &\lambda^4 \\
*  & 1 \lambda^3  & \lambda^2\\
*  &           *   & 1 \\
\end{pmatrix}
\;,
&
Y_\mathrm{e}=
\begin{pmatrix}
        0  &-3\lambda^4 & \lambda \\
*  & -3\lambda^3  &1\\
*  &         0  & 1 \\
\end{pmatrix}\;,
&
m_\mathrm{LL} \approx \begin{pmatrix}
0  & m    &  0 \\
m  & 0 &0  \\
0  &0 &0 \\
\end{pmatrix}
 .
 \\
\end{array}
\end{eqnarray}
Some comments are in order: Since we have chosen $(Y_\mathrm{e})_{32}$ to be
very small (say $\ll \lambda^4$), $\theta_{12}^\mathrm{e}$ is given by
$\theta_{12}^\mathrm{e} \approx (Y_\mathrm{e})_{12}/(Y_\mathrm{e})_{22} \approx
\lambda$. The rotation by $R_{13}^\mathrm{e}$ then has to generate a $0$-entry in
$(Y_\mathrm{e})_{13}$, which typically leads to
$\theta_{13}^\mathrm{e} =\mathscr{O}(\lambda)$, however
it can well be smaller due to cancellations. Of course,
$\theta_{23}^\mathrm{e}$ close to maximal follows from the lopsided structure of
$Y_\mathrm{e}$.
The operators for the (1,2)- and (2,2)-entries in $Y_\mathrm{e}$ and
$Y_\mathrm{d}$ have to be chosen such that we obtain Clebsch factors $-3$ for
$(Y_\mathrm{e})_{12}/(Y_\mathrm{d})_{12}$ and
 for $(Y_\mathrm{e})_{22}/(Y_\mathrm{d})_{22}$.
As discussed above, for a lopsided $Y_\mathrm{e}$ the equal 
Clebsch factors are required for realizing the desired relation $\theta_{12} \approx \tfrac{\pi}{4} -
\theta_\mathrm{C}$ at $M_\mathrm{X}$. For the quark sector, models of this type
predict $m_\mu/m_s = 3 \cos(\theta_{23}^\mathrm{e})$ at $M_{\mathrm{X}}$.
 In contrast to
 the model in Sec.~\ref{Sec:ModelA} with large $\theta_{23}$ originating from
 the neutrino sector, such lopsided QLC models do not firmly
 predict a large $\theta_{13}$ equal to the Cabibbo angle.

\providecommand{\bysame}{\leavevmode\hbox to3em{\hrulefill}\thinspace}


\begin{thebibliography}{1}

\bibitem{King:2003jb} 
see e.g.:
R.~N.~Mohapatra and P.~B.~Pal,
  [World Sci.\ Lect.\ Notes Phys.\  {\bf 72} (2004) 1].
S.~F.~King,
Rept.\ Prog.\ Phys.\  {\bf 67} (2004) 107, hep-ph/0310204.

\bibitem{King:2000ce}
S.~F. King and N.~N. Singh, 
Nucl. Phys. \textbf{B596} (2001), 81--98,  {hep-ph/0007243}.



\bibitem{King:2002nf}
S.~F. King, 
JHEP \textbf{09} (2002), 011,
  {hep-ph/0204360}.

\bibitem{Frampton:2004ud}
  P.~H.~Frampton, S.~T.~Petcov and W.~Rodejohann,
  Nucl.\ Phys.\ B {\bf 687} (2004) 31, hep-ph/0401206.


\bibitem{GST}
  R. Gatto, G. Sartori and M. Tonin, Phys. Lett. B {\bf 28} (1968) 128.

\bibitem{Georgi:1979df}
  H.~Georgi and C.~Jarlskog,
  Phys.\ Lett.\ B {\bf 86} (1979) 297.
  
\bibitem{Raidal:2004iw}
  M.~Raidal,
  Phys.\ Rev.\ Lett.\  {\bf 93} (2004) 161801, hep-ph/0404046.

\bibitem{Minakata:2004xt}
  H.~Minakata and A.~Y.~Smirnov,
  Phys.\ Rev.\ D {\bf 70}, 073009 (2004), hep-ph/0405088.

\bibitem{QLCliterature} For recent papers related to QLC, see e.g.: 
  J.~Ferrandis and S.~Pakvasa,
  Phys.\ Rev.\ D {\bf 71} (2005) 033004, 
  hep-ph/0412038;
  S.~K.~Kang, C.~S.~Kim and J.~Lee,
  hep-ph/0501029; 
  N.~Li and B.~Q.~Ma,
  hep-ph/0501226;
  K.~Cheung, S.~K.~Kang, C.~S.~Kim and J.~Lee,
  hep-ph/0503122;
  D.~Falcone,
  hep-ph/0503197; 
  Z.~z.~Xing,
  hep-ph/0503200; 
  A.~Datta, L.~Everett and P.~Ramond,
  hep-ph/0503222.


\bibitem{ps} J. C. Pati and A. Salam, Phys. Rev. {\bf D 10}, 270 (1974).

\bibitem{mohfram} 
  P.~H.~Frampton and R.~N.~Mohapatra,
  JHEP {\bf 0501}, 025 (2005), hep-ph/0407139.
  
\bibitem{Roberts:2001zy}
R.~G. Roberts, A.~Romanino, G.~G. Ross, and L.~Velasco-Sevilla, 
Nucl. Phys. \textbf{B615} (2001), 358--384,
  {hep-ph/0104088}.

\bibitem{Allanach:1996hz}
B.~C. Allanach, S.~F. King, G.~K. Leontaris, and S.~Lola, 
Phys. Rev. \textbf{D56} (1997), 2632--2655,
  {hep-ph/9610517}.

\bibitem{King:1998nv}
S.~F. King and M.~Oliveira, 
Phys. Rev. \textbf{D60} (1999), 035003,  {hep-ph/9804283}.



\bibitem{emutau}  S. Petcov, Phys. Lett.{\bf B
10}, 245 (1982); R. Barbieri, L. Hall, D. Smith, A. Strumia and N.
Weiner, JHEP {\bf 9812}, 017 (1998); A. Joshipura and S. Rindani,
Eur.Phys.J. {\bf C14}, 85 (2000); R. N. Mohapatra, A.
Perez-Lorenzana, C. A. de S. Pires, Phys. Lett. {\bf B474}, 355
(2000); T. Kitabayashi and M. Yasue, Phys. Rev. {\bf D 63},
095002 (2001); Phys. Lett. {\bf B 508}, 85 (2001);
hep-ph/0110303;  L. Lavoura, Phys. Rev. D 62, 093011 (2000);
 W. Grimus and L. Lavoura, Phys. Rev. D 62, 093012 (2000);
 J. High Energy Phys. 09, 007 (2000); J. High Energy Phys. 07, 045 (2001);
R. N. Mohapatra, hep-ph/ 0107274; Phys. Rev. {\bf D 64}, 091301
(2001). K. S. Babu and R. N. Mohapatra, Phys. Lett. {\bf B 532},
77 (2002); H. S. Goh, R. N. Mohapatra and S.-P. Ng,
hep-ph/0205131; Phys. Lett. {\bf B542}, 116 (2002)); Duane A.
Dicus, Hong-Jian He, John N. Ng, Phys. Lett. {\bf B 536}, 83
(2002); Q. Shafi and Z. Tavartkiladze, Phys. Lett. {\bf B 482},
1451 (2000).

\bibitem{Antusch:2005gp}
S.~Antusch, J.~Kersten, M.~Lindner, M.~Ratz, and M.~A. Schmidt, 
  {hep-ph/0501272}.


\bibitem{Antusch:2003kp}
  S.~Antusch, J.~Kersten, M.~Lindner and M.~Ratz,
  Nucl.\ Phys.\ B {\bf 674} (2003) 401, 
  hep-ph/0305273.
  
  
\end{thebibliography}
\end{document}